\documentclass[prl,twocolumn,epsfig,rotate]{revtex4}
\usepackage{graphicx}
\usepackage{amsmath}
\usepackage{dcolumn}
\usepackage{epsfig}
\usepackage{bm}
\usepackage{mathrsfs}
\begin{document}

\title{Observing Golden Mean Universality Class in the Scaling of Thermal Transport}

\author{Daxing Xiong}
\email{phyxiongdx@fzu.edu.cn}
\affiliation{Department of Physics,
Fuzhou University, Fuzhou 350108, Fujian, China}

\begin{abstract}
We address the issue of whether the golden mean
$\left[\psi=(\sqrt{5}+1)/2 \simeq 1.618\right]$ universality class,
as predicted by several theoretical models, can be observed in the dynamical scaling of thermal transport. Remarkably, we show estimate with unprecedented precision, that $\psi$ appears to be the scaling
exponent of heat mode correlation in a purely quartic
anharmonic chain. This observation seems somewhat deviation from the
previous expectation and we explain it by the unusual slow decay of
the cross-correlation between heat and sound modes. Whenever the
cubic anharmonicity is included, this cross-correlation is gradually
died out and another universality class with scaling exponent
$\gamma=5/3$, as commonly predicted by theories, seems
recovered. However, this recovery is accompanied by two interesting phase transition processes characterized by a change of symmetry of the potential and a clear
variation of the dynamic structure factor, respectively. Due to these transitions, an additional exponent close to $\gamma \simeq 1.580$ emerges. All these evidences
suggest that, to gain a full prediction of the scaling of
thermal transport, more ingredients should be taken into account.
\end{abstract}
\maketitle

\emph{Introduction.}---Despite decades of intensive studies~\cite{Report2016,Report2003,Report2008,Lebowitz1967,Casati1984,Lepri1997,Zhao1998,Prosen2000,Grassberger2002,Narayan2002,Casati2004,Politi2005,LeeDadswell2005,Thesis1, Olla2006,Zhao2006,MCT2007,Dhar2007,LeiWang2011,Denisov2011,Beijeren2012,Spohn2013,Peter2014,Casati2017,Olla2017,Satio2017}, our understanding of thermal transport in one-dimensional (1D) systems is still scarce. Macroscopically, such transport is described by an empirical law, i.e., the Fourier's law:
$\emph{J}=- \kappa \nabla T$ with $\emph{J}$~~the heat current, $ \nabla T$ the spatial temperature gradient, and $\kappa$ a material \emph{constant} named as thermal conductivity. Nevertheless, now it has been generally realized that Fourier's law is not always valid; instead, an anomalous transport will be shown in most cases~\cite{Report2016,Report2003,Report2008}. In particular, in the momentum-conserving systems which are of particular interest, this anomaly is mainly characterized by~\cite{Note1}: $\alpha$ describing the divergence of $\kappa$ with increasing space size $L$ as $\kappa \sim L^{\alpha}$ with $0<\alpha\leq1$ (normal transport, $\alpha=0$)~\cite{Lepri1997, Prosen2000,Grassberger2002,Narayan2002,Casati2004,LeeDadswell2005,Thesis1,Olla2006,Dhar2007,MCT2007,LeiWang2011,Casati2017,Satio2017}, and $\gamma$ giving the space($m$)-time($t$) scaling of energy/heat correlation $\rho_{{E}/{Q}} (m,t)$ by $t^{-\frac{1}{\gamma}} \rho_{{E}/{Q}} (\frac{m}{ t^{1/\gamma}},t)$ with $1\leq \gamma < 2$ (normal transport, $\gamma=2$)~\cite{Politi2005,Zhao2006, Denisov2011,Beijeren2012,Spohn2013,Peter2014}. Based on L\'{e}vy walk assumption~\cite{Politi2005,Report2015}, there is a generic formula $\alpha=2-\gamma$ linking $\alpha$ and $\gamma$, while currently, the universality classes of both exponents remain unclear.

Early studies focused on $\alpha$. Based on mode coupling theory~\cite{Lepri1997,MCT2007}, hydrodynamics renormalization group approach~\cite{Narayan2002}, solvable model of stochastic dynamics~\cite{Olla2006}, and numerical investigations~\cite{Lepri1997,Grassberger2002,MCT2007,Dhar2007,LeiWang2011}, three universality classes, $\alpha=2/5$, $\alpha=1/3$, and $\alpha=1/2$, at that time, are the most common ones. However, such universality classification was doubted~\cite{Xiong2012,Hurtado2016}, and later, a mode cascade theory (MCT)~\cite{LeeDadswell2005,Thesis1} took the peculiar coupling between energy and momentum transport into account and suggested that, generally, $\alpha$ belongs to a Fibonacci universality sequence and will finally converge to one kind of golden mean values $\alpha^{*}=(3-\sqrt{5})/2 \simeq 0.382$.

The scaling exponent $\gamma$ is of interest since heat transport is related to energy/heat diffusion process. Several pioneering studies~\cite{Politi2005,Zhao2006,Denisov2011} employed energy correlation and obtained $\gamma=5/3$. Later, since 2012 additional insights have been gained from the tool of nonlinear fluctuating hydrodynamic theory (NFHT)~\cite{Beijeren2012,Spohn2013}, which considered the full three normal modes (one heat mode and two sound modes) correlations. Based on this, van Beijeren~\cite{Beijeren2012} first predicted $\gamma=5/3$ universality class for heat mode correlation in general Hamiltonian dynamics with three conservation laws. Such universality class was subsequently demonstrated in Fermi-Pasta-Ulam (FPU) chains with asymmetric potentials~\cite{Spohn2013}, and generalized to an arbitrary anharmonic chain but with another universality class $\gamma=3/2$ (for symmetric potentials under zero pressure) reported~\cite{Spohn2014}. Recently, two research groups studied the systems when the sound modes are absent~\cite{Olla2017,Satio2017}. Based on explicitly solvable models of stochastic dynamics~\cite{Olla2006}, they observed normal~\cite{Olla2017} and anomalous (with new exponents reported)~\cite{Satio2017} transport, respectively.

Since NFHT addresses the hydrodynamic description for conserved quantities, quite recently it has been adopted to study the relevant dynamical structure function in general transport processes far away from equilibrium, from which several similar scaling exponents as $\gamma$~\cite{Note-r} have been extracted and their universality classes have been discussed~\cite{Popkov2014,Popkov2015,Spohn2015,PopkovPNAS,Popkov2016}. In particular, in the presence of several conserved quantities, a Fibonacci universality sequence of $\gamma$ converging to another kind of golden mean values $\psi=(\sqrt{5}+1)/2 \simeq 1.618$~\cite{PopkovPNAS} was predicted. Subsequent demonstrations showed that this $\gamma=\psi$ universality class can also appear in the case of two conservation laws~\cite{Popkov2015,Spohn2015,Popkov2016}. Such progress in turn provided renewed insight into the scaling of 1D thermal transport, which led Spohn~\cite{SpohnArxiv} to reformulate the universality classification to include the new classification $\gamma=\psi$ and further argue that this is induced by the cross-coupling between different modes. Nevertheless, still at present, the $\gamma=\psi$ classification has not yet been reported in any numerical studies of thermal transport~\cite{Note2}.

In this Letter, we therefore address the question: whether $\psi$ can be observed in the scaling of thermal transport in an anharmonic chain system. We will try to find such a system from \emph{numerical} perspective.
Remarkably, our following estimate of a purely quartic anharmonic chain just gives $\gamma=\psi$ with high precision.
Such system's Hamiltonian is
\begin{equation} \label{Hamiltonian}
H= \sum_{m=1}^{L} p_{m}^2/2 + V(r_{m+1}-r_m)
\end{equation}
with $p_m$ the $m$th particle's momentum, $r_{m}$ its displacement from equilibrium position, and $V(\xi)= \xi^4/4$. This Hamiltonian has a scaling property, i.e., $H/T=H'/T'$ under $p_m=\left( T/T' \right)^{\frac{1}{2}} p_{m}'$ and $r_m=\left( T/T' \right)^{\frac{1}{4}} r_{m}'$, resulting in a temperature-independent thermal transport~\cite{Zhao2005}. Therefore, the system was conjectured to be an ideal, clean anharmonic subject~\cite{LeeDadswell2005,Thesis1}. However, the existing studies seem not well capturing such peculiar system's transport. Early prediction claimed $\alpha=2/5$ but the relevant numerical studies only suggested $\alpha \simeq 0.37$ and at that time it was thought to be close enough to the prediction~\cite{Lepri1998}. The quite recent predictions gave $\alpha=1/2$~\cite{LeeDadswell2005,Thesis1} and $\gamma=3/2$ (since potential is symmetric)~\cite{Spohn2014}, but the convincing numerical results of $\alpha$ only indicated $\alpha \simeq 0.38$~\cite{LeeDadswell2005,Thesis1} and there is not any numerical estimate of $\gamma$ reported. It is thus very strange why such a clean system can result in so confused conclusions. In view of this obscure, and inspired by the new insight of universality classification~\cite{Popkov2014, Popkov2015,Spohn2015,PopkovPNAS,Popkov2016,SpohnArxiv}, it is surely necessary to check this clean subject's universality class with more precise estimate, in particular for $\gamma$.

\emph{Golden mean universality class for $\gamma$.}---To obtain an accurate estimate of $\gamma$, we employ the correlation function of heat energy fluctuation $\rho_{Q}(m,t)=\frac{\langle \Delta Q_{i+m}(t) \Delta Q_{i}(0) \rangle}{\langle \Delta Q_{i}(0) \Delta Q_{i}(0) \rangle}$~\cite{Zhao2013}, where $\langle \cdot \rangle$ denotes the spatiotemporal average; $Q_i(t) \equiv E_i(t)-\frac{(\langle E \rangle + \langle F\rangle) g_i(t)}{\langle g \rangle}$~\cite{Beijeren2012} is the heat energy density at a coarse-grained location $i$ and time $t$ with $g_i$, $E_i$, and $F_i$ the particle, energy, and pressure densities, respectively; $\Delta Q_i(t) \equiv Q_i(t)-\langle Q \rangle$ is its fluctuation. Such a coarse-grained $\rho_{Q}(m,t)$ just corresponds to the heat mode correlation in hydrodynamics~\cite{Beijeren2012,Zhao2013}, and thus $\rho_{Q}(m,t)$ can be adopted to accurately extract $\gamma$. The simulations are performed under an appropriate temperature $T=0.5$ by the usual molecular dynamics simulations and we refer them to~\cite{Note3}. One point we remind that we set both the equilibrium distance
between particles as well as the lattice constant to unity. This makes
the number of particles equal to the system size $L$, and therefore, for systems with symmetric potential, the
average pressure $\langle F \rangle \equiv 0$.
\begin{figure}
\begin{centering}
\vspace{-.6cm} \includegraphics[width=8.8cm]{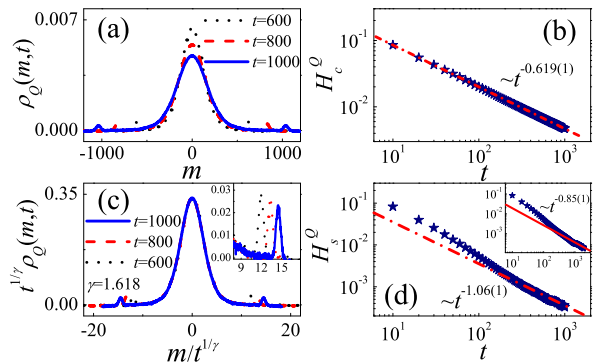} \vspace{-.8cm}
\caption{\label{Fig1} The purely quartic anharmonic chain: (a) [(c)] $\rho_{Q}(m,t)$ [Rescaled $\rho_{Q}(m,t)$] for three typical long times; (b)[(d)] The height $H_c^{Q}$ [$H_s^{Q}$] of the central (side) peak of $\rho_{Q}(m,t)$ vs $t$. The inset in (c) is a zoom for the right side peaks; the inset in (d) shows the same plot as (d) but with a longer time $t=1900$.} \vspace{-.3cm}
\end{centering}
\end{figure}

Figure~\ref{Fig1}(a) depicts the measured $\rho_{Q}(m,t)$ for three long times. As expected, there are three L\'{e}vy walk-like profiles~\cite{Denisov2011,Report2015} with a central peak along with two side peaks. With
these profiles and based on L\'{e}vy walk theory~\cite{Denisov2011,Report2015}, one then can perform a scaling analysis~\cite{Denisov2011,Report2015}
\begin{equation} \label{Scaling}
t^{1/\gamma} \rho(m,t) \simeq \rho(t^{-1/\gamma}m,t),
\end{equation}
to infer $\gamma$. In practice, $\gamma$ is extracted from the time scaling of the height $H_c^{Q}$ of the central peak by $H_c^{Q} \sim t^{-1/\gamma}$. Figs.~\ref{Fig1}(b) and (c) present both results. To our own surprise, a perfect scaling with $\gamma$ nearly exactly located at $\psi$ can be clearly detected. In particular, Fig.~\ref{Fig1}(b) does show a faultless scaling for all the considered times, which is reasonable since the focused system is an ideal, clean anharmonic subject~\cite{LeeDadswell2005,Thesis1}.

\emph{Cross-correlation between heat and sound modes.}---Clearly, our above estimate is deviation from the prediction $\gamma=3/2$~\cite{Spohn2014}. This leads us to carefully consider the cross-correlation between heat and sound modes, for which we explore the time scaling of the height $H_s^{Q}$ of $\rho_Q(m,t)$'s side peaks~\cite{Dhar2014}. As the scaling $H_s^{Q} \sim t^{-1.06 \pm 0.01}$ [see Fig.~\ref{Fig1}(d)] is now not so faultless, we have also performed an estimate with a nearly twice longer time. Undoubtedly, $\gamma=\psi$ for central peak is still recovered (no shown), but the new estimate suggests $H_s^{Q} \sim t^{-0.85 \pm 0.01}$ and with a trend towards lower values [see the inset of Fig.~\ref{Fig1}(d)], which is almost comparable to the scaling of $H_c^{Q}$. This implies that, as time increases, the decay of the cross-correlation will become slower and slower.

With such unusual evidence, we then turn to NFHT, and the point of deviation can then be understandable.
To obtain the two universality classes of $\gamma=5/3$ and $\gamma=3/2$, NFHT requires to use the decoupling hypothesis (in a asymptotic long time)~\cite{Spohn2014}. That is why, in the prediction the heat mode correlation usually does not show additional side peaks, or in other words, the side peaks are assumed to decay very quickly. However, our estimate here shows that, for the focused purely quartic anharmonic chain, this is not always the case. In fact, it is just the slowly decaying cross-correlation between different modes resulting in the new classification $\gamma=\psi$, which is basically consistent with the recent theoretical conjecture of NFHT~\cite{Popkov2015,Spohn2015,PopkovPNAS,Popkov2016,SpohnArxiv}.

In fact, the purely quartic anharmonic system has not yet been analyzed numerically in the NFHT scheme~\cite{Note-r2}. In view of this, we have measured the relevant heat and sound modes correlations following the NFHT standard (see~\cite{Note3}). The heat mode correlations for different times show similar profiles as our estimate in Fig.~\ref{Fig1}, and the scaling exponent $\gamma$ for central peak exhibits deviation from $\gamma=3/2$ as well. Interestingly, the scaling of the side peaks does display a similar slow decay. Besides, certain correlation between two sound modes can also be detected. Clearly, this last point has not yet been addressed by the theory. All these evidences seem not contradicting our above explanation.
\begin{figure}
\begin{centering}
\vspace{-.6cm}\includegraphics[width=7cm]{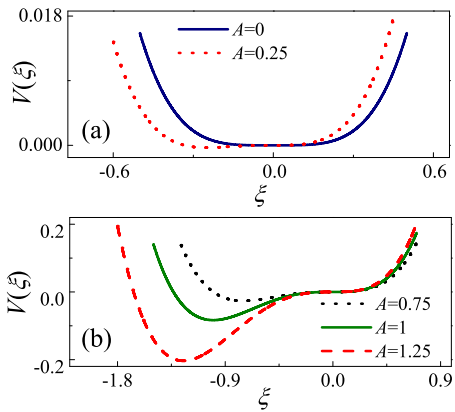} \vspace{-.5cm}
\caption{\label{Fig2} Potentials of cubic-plus-quartic anharmonic chain for different $A$.}
\vspace{-.3cm}
\end{centering}
\end{figure}

\emph{Discussion of $\alpha$.}---$\gamma=\psi$ implies $\alpha=\alpha^{*} \simeq 0.382$, if the L\'{e}vy walk assumption~\cite{Politi2005,Report2015} is valid. Fortunately, Fig.~\ref{Fig1} seems to support such an assumption and thus $\alpha=2-\gamma$ appears valid. Based on this, even though here $\gamma$ is our main focus, we are able to discuss $\alpha$. For the early prediction $\alpha=2/5$~\cite{Lepri1998}, as mentioned, this deviation is understandable, since at that time, no additional insight of $\gamma$ had been gained, and $\alpha=2/5$ was already so close to $\alpha=\alpha^{*}$. In fact, such renewed understanding also took place by a more accurate estimate of a similar scaling exponent in $2+1$ dimensional Kardar-Parisi-Zhang equation~\cite{Miss2015}.

For the prediction $\alpha=1/2$, we remind that MCT~\cite{LeeDadswell2005,Thesis1} resorts to one key hypothesis, i.e., the coupling between energy and momentum transport, and due to this coupling, very long-time estimate thus requires. While when applying to the purely quartic anharmonic system, such coupling was conjectured to be particularly simple: energy transport depends only on momentum transport~\cite{LeeDadswell2005,Thesis1}. This latter conjecture surely indicates possible coupling between heat and sound modes, coincident with our above explanation. Viewing this, we have performed a large-scale estimate (with system size up to $L=163840$) for $\alpha$ (see~\cite{Note3}). While a scaling with $\alpha=0.388 \pm 0.003$ seems well fitted, in turn implying $\gamma=\psi$. Therefore, whether the prediction $\alpha=1/2$ would be observed for a yet larger $L$ remains open~\cite{LeeDadswell2005,Thesis1,LeiWang2011}.

In spite of the above unclear point, we suggest that, inspired by MCT~\cite{LeeDadswell2005,Thesis1} and combining the results of Fig.~2 in~\cite{Xiong2016}, one might understand $\gamma=\psi$ by a similarly anharmonicity induced cascade process with a convergence $\gamma=\psi$, since the purely quartic anharmonic chain is a highly anharmonic limit of the FPU model considered in~\cite{Xiong2016}.

\emph{Introducing cubic anharmonicity to destroy the cross-correlation.}---We next ask the question: whenever such cross-correlation disappears, which universality class will be recovered. Toward this aim, we include the cubic anharmonicity to study a cubic-plus-quartic anharmonic chain~\cite{Lee-Dadswell2008}, i.e., system with Hamiltonian~\eqref{Hamiltonian} and $V(\xi)= {A\xi^3}/3+\xi^4/4$. Here, $A$ ($A > 0$) controls the comparative strength of the cubic to quartic anharmonicity. It causes the potential no longer being symmetric (see several plots in~Fig.~\ref{Fig2}). NFHT then predicted that such system follows $\gamma=5/3$~\cite{Beijeren2012,Spohn2014}, while at present no further numerical estimate has been presented.

\begin{figure}
\begin{centering}
\vspace{-.6cm} \includegraphics[width=8.8cm]{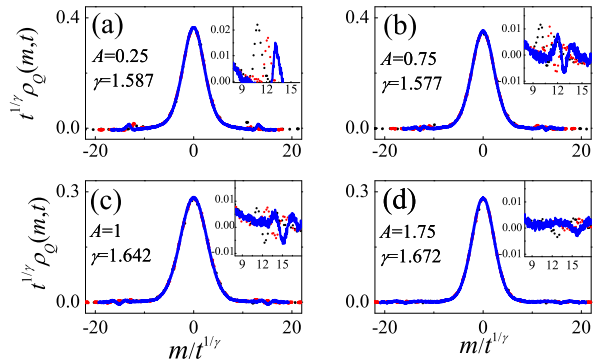} \vspace{-.8cm}
\caption{\label{Fig3} The cubic-plus-quartic anharmonic chains: Rescaled $\rho_{Q}(m,t)$ ($T=0.5$) for three typical long times (here and below, the same as Fig.~\ref{Fig1}) for (a) $A=0.25$, (b) $A=0.75$, (c) $A=1$, and (d) $A=1.75$, with measured average pressures $\langle F \rangle \simeq -0.176$, $\langle F \rangle \simeq -0.543$, $\langle F \rangle \simeq -0.744$, and $\langle F \rangle \simeq -1.503$, respectively. The insets are used
as a zoom for the side parts (right).} \vspace{-.3cm}
\end{centering}
\end{figure}

Figure~\ref{Fig3} depicts the measured rescaled $\rho_{Q}(m,t)$. Indeed, with the increase of $A$, $\rho_{Q}(m,t)$'s side peaks are distorted first [see the insets of~Figs.~\ref{Fig3}(a)-(c)], and they seem finally disappeared, as long as $A$ is large enough [see for example the inset of~Fig.~\ref{Fig3}(d)]. This indicates that including cubic anharmonicity does destroy the cross-correlation between heat and sound modes, and eventually, a scaling exponent $\gamma \simeq 1.672$ [see~Fig.~\ref{Fig3}(d)], which is close to the prediction $\gamma=5/3 \simeq 1.667$, can be observed.

\begin{figure}
\begin{centering}
\vspace{-.6cm}\includegraphics[width=7cm]{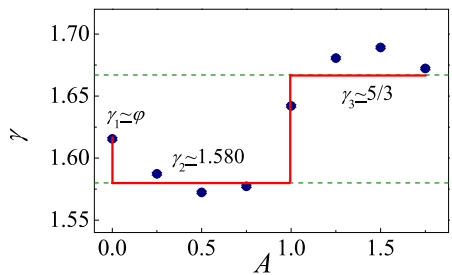} \vspace{-.5cm}
\caption{\label{Fig4} The cubic-plus-quartic anharmonic chains: $\gamma$ vs $A$. The error bars are smaller than the symbol size.} \vspace{-.3cm}
\end{centering}
\end{figure}

\emph{Three observed universality classes.}---Through a detailed examination of $\gamma$ versus $A$ (Fig.~\ref{Fig4}), one might roughly identify three universality classes, i.e., (I) $\gamma_1=\psi$ ($A=0$), (II) $\gamma_2 \simeq 1.580$ ($0<A<1$), and (III) $\gamma_3=5/3$ ($A \geq 1$). As mentioned, $\gamma_1$ and $\gamma_3$ are resulted from a nonvanishing and vanishing cross-correlation between heat and sound modes, respectively; and thus $\gamma_2$ can be understood by a transition process where a distorted cross-correlation arises. Further examination shows that this distortion is induced by an additional negative correlation between heat and sound modes at the location of $\rho_Q(m,t)$'s side peaks (see~\cite{Note3}). More-interestingly, the second transition process seems discontinous (Fig.~\ref{Fig4}), suggesting a feature similarly to that of structure phase transition~\cite{Casati2017,Xiong2016}.

\emph{Phase transitions.}---We employ Fig.~\ref{Fig2}(a) to phenomenologically understand the transition from (I) to (II). This is mainly induced by the potential changed from a symmetric to an asymmetric type, which causes certain negative correlation between heat and sound modes (as mentioned above, see~\cite{Note3}). Clearly, such slight variation of correlations cannot be detected by only measuring $\alpha$~\cite{Zhao2012,Baowen2013,Dhar2014JPhys,Zhao2016}. Then if the formula $\alpha=2-\gamma$~\cite{Politi2005,Report2015} is still valid, it causes the scaling exponent $\alpha$ increased from $\alpha=\alpha^{*}$ to $\alpha \simeq 0.420$. Such $\alpha$ value appearing in the cubic-plus-quartic anharmonic chain under appropriate $A$ seems somewhat distinct from those in other typical systems with asymmetric potentials~\cite{Zhao2012,Baowen2013,Dhar2014JPhys,Zhao2016}.

For the transition from (II) to (III), we regard it as a temperature/anharmonicity induced structure phase transition. Let us now turn to Fig.~\ref{Fig2}(b), which shows potentials around the transition point $A_{\rm cr}=1$. From Fig.~\ref{Fig2}(b), for an arbitrary $A$, there should be a corresponding critical temperature $T_{\rm cr}$ (and vice versa), below which particles can only be trapped by the left well, whereas above $T_{\rm cr}$, particles are free but trapped by the whole big well. Our results of Figs.~\ref{Fig3} and~\ref{Fig4} just suggest that, for $T=0.5$, $A_{\rm cr}=1$ correspondingly.

To further verify the second conjecture, we finally estimate the system's dynamic structure factor, i.e., the particles' density-density correlation function, defined by $\rho_{g}(m,t)=\frac{\langle \Delta g_{i+m}(t) \Delta g_{i}(0) \rangle}{\langle \Delta g_{i}(0) \Delta g_{i}(0) \rangle}$~\cite{Beijeren2012}. Here, as mentioned, $g_i(t)$ is the particle density and thus $\Delta g_{i}(t)=g_i(t)- \langle g \rangle$ is its fluctuation. Simulation of $\rho_{g}(m,t)$ is also similarly to that of $\rho_{Q}(m,t)$~\cite{Note3}. In general, $\rho_{g}(m,t)$ is conjectured to exhibit one central Rayleigh peak and two side Brillouin peaks, due to the anomalous transport~\cite{Beijeren2012}.

\begin{figure}
\begin{centering}
\vspace{-.6cm} \includegraphics[width=8.8cm]{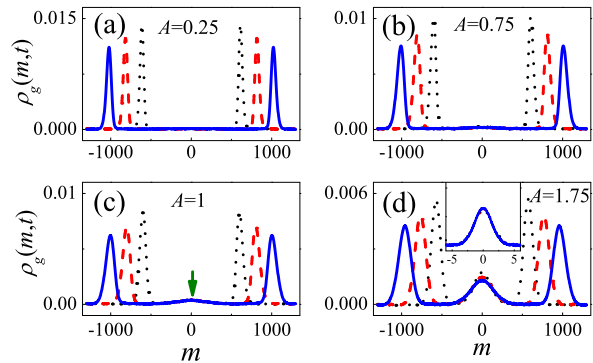} \vspace{-.8cm}
\caption{\label{Fig5} The cubic-plus-quartic anharmonic chains: $\rho_{g}(m,t)$ ($T=0.5$) for three typical long times for (a) $A=0.25$, (b) $A=0.75$, (c) $A=1$, and (d) $A=1.75$. The inset in (d) shows the rescaled central parts of $\rho_{g}(m,t)$ under formula~\eqref{Scaling} with $\gamma=5/3$.} \vspace{-.3cm}
\end{centering}
\end{figure}

Figure~\ref{Fig5} depicts the measured $\rho_{g}(m,t)$ for several typical $A$. As expected, around $A_{\rm cr}=1$, there is a clear variation of $\rho_{g}(m,t)$, i.e., below $A_{\rm cr}=1$, the central Rayleigh peak is absent, while above $A_{\rm cr}=1$, this peak starts to emerge, due to the $\gamma=5/3$ [see the inset of Fig.~\ref{Fig5}(d)] universality class~\cite{Beijeren2012}. Such a variation of $\rho_{g}(m,t)$ does support our conjecture of structure phase transition.

\emph{Conclusion.}---We have presented estimate with high precision and shown that, the golden mean universality class $\gamma=\psi$ can appear in the scaling of heat mode correlation in a purely quartic anharmonic chain. The key reason for this universality is that, there exists an unusual slow decay of the cross-correlation between heat and sound modes. This understanding is basically consistent with the recent argument of NFHT~\cite{Popkov2015,Spohn2015,PopkovPNAS,Popkov2016,SpohnArxiv}. Including cubic anharmonicity can destroy this cross-correlation rapidly, and eventually, $\gamma$ will be changed into $\gamma=5/3$ universality class, as expected by the theory~\cite{Spohn2014}. However, between $\gamma=\psi$ and $\gamma=5/3$, there are two interesting phase transition processes where an additional exponent $\gamma \simeq 1.580$ seems to appear. Further examination shows that, the first transition is induced by the change of symmetry of the potential resulting in a distorted cross-correlation; the second transition is due to the anharmoncity induced structure phase transition eventually destroying the cross-correlation. Clearly, such two transitions suggest some new ingredients that should be carefully considered in further examining the scaling of thermal transport. Surely, all these rich evidences have not yet been covered by present theories, and currently they can only resort to numerical studies. Hopefully, the observations can provide insight and further advance the theories.

\begin{acknowledgments}
D.X. thanks the referees for many useful suggestions, and appreciates Prof. H. Spohn for helpful discussion. This work was supported by the NNSF (Grant No. 11575046) of China; the NSF (Grant No. 2017J06002) of Fujian provice; the training plan for Distinguished Young researchers of Fujian provincial department
of education; the Qishan scholar research fund of Fuzhou university.
\end{acknowledgments}

\pagebreak
\clearpage
\begin{center}
\textbf{\large Supplementary Material for
`Observing Golden Mean Universality Class in the Scaling of Thermal Transport'}
\end{center}
\setcounter{equation}{0} \setcounter{figure}{0}
\setcounter{table}{0} \setcounter{page}{1} \makeatletter
\renewcommand{\theequation}{S\arabic{equation}}
\renewcommand{\thefigure}{S\arabic{figure}}
\renewcommand{\bibnumfmt}[1]{[S#1]}
\renewcommand{\citenumfont}[1]{S#1}
\emph{1. Simulation details.}---
In the main text, $\rho_Q(m,t)$ is defined by
\begin{equation}
\rho_{Q}(m,t)=\frac{\langle \Delta Q_{i+m}(t) \Delta Q_{i}(0) \rangle}{\langle \Delta Q_{i}(0) \Delta Q_{i}(0) \rangle}.
\end{equation}
Herein $Q_i(t)$ is defined as the heat density within a finite volume (bin i) at time $t$ whose expression is~\cite{S_Forster,S_Liquid,S_Beijeren2012}
\begin{equation} \label{Q}
Q_i(t) \equiv E_i(t)-\frac{(\langle E \rangle + \langle F\rangle) g_i(t)}{\langle g \rangle}.
\end{equation}
This expression is derived from basic thermodynamics in conventional hydrodynamic theory~\cite{S_Forster,S_Liquid,S_Beijeren2012} and therein, $g_i$, $E_i$, and $F_i$ are the number of particles, the energy, and the pressure within the bin, respectively; $\langle \cdot \rangle$ represents the spatiotemporal average.

To simulate $Q_i(t)$, we first divide the chain into several equivalent bins. In each bin, we then calculate $g_i$, $E_i$ and $F_i$ within the bin. Finally, $Q_i(t)$ can be computed according to Eq.~\eqref{Q} and its fluctuation then is $\Delta Q_i(t)= Q_i(t)-\langle Q \rangle$.

To simulate $\rho_Q(m,t)$ and $\rho_g(m,t)$, in most cases we consider a chain with $L=4001$
particles, which allows an initial heat or particle fluctuations located at the
center to spread out a lag time at least up to $t=1000$. Note that if one needs to simulate a longer time, a longer size is required. We apply periodic
boundary conditions and fix the number of bins to be $(L-1)/2$. We
use the stochastic Langevin heat baths~\cite{S_Lepri_Report,S_Dhar_Report} to thermalize the system and
to prepare a canonical equilibrium state to the focused temperature
$T=0.5$. Under these setups, we employ the
Runge-Kutta algorithm of seventh to eighth order with a time step of
$h=0.05$ to evolve the system. Each canonical equilibrium system is
prepared by evolving the system for a long enough time ($>10^7$ time
units) from properly assigned initial random states. Finally, we use
ensembles of about $8\times10^9$ data points to compute both
correlation functions.

\emph{2. Correlation functions under the NFHT scheme.}---Instead of using the three locally conserved quantities: particle number $g_i$, momentum $p_i$, and energy $E_i$ in a coarse-grained bin description, in conventional hydrodynamic theory~\cite{S_Forster,S_Liquid,S_Beijeren2012}, NFHT~\cite{S_Spohn2013,S_Spohn2014,S_Dhar2014} adopts the stretch $s_m=r_{m+1}-r_m$, momentum $p_m$, and energy $E_m$ for conserved quantities for each labelling particle. With these conserved quantities, one then can consider small fluctuations:
$u_1(m,t)=s_m(t)-\langle s \rangle$, $u_2(m,t)=p_m(t)-\langle p \rangle$, and $u_3(m,t)=E_m(t)-\langle E \rangle$, and construct the corresponding conserved field $\vec{u}=(u_1,u_2,u_3)$ for the fluctuating hydrodynamic equations.

The quantities of interest are the equilibrium correlations of $\vec{u}=(u_1,u_2,u_3)$ with elements:
\begin{equation} \label{CC}
C_{\mu\nu} (m,t)=\langle u_{\mu}(m,t) u_{\nu}(0,0) \rangle,
\end{equation}
where $\mu, \nu \equiv 1, 2, 3$. NFHT suggests that one can switch to normal modes correlations through the transformation $(\varphi_{-1},\varphi_0,\varphi_1)=\vec{\varphi}=R \vec{u}$, where $\varphi_{\pm 1}$ represent two sound modes traveling at sound speed in opposite directions, $\varphi_{0}$ corresponds to the stationary heat mode, and $R$ is a matrix known analytically in Eq.~(8.13) of~\cite{S_Spohn2014}.

With this transformation, one then can move to the normal modes correlation functions:
\begin{equation} \label{CC}
C_{\sigma \sigma'}(m,t)=\langle \varphi_{\sigma}(m,t) \varphi_{\sigma'}(0,0) \rangle,
\end{equation}
where $\sigma, \sigma' \equiv -1, 0, 1$.

\begin{figure*}
\begin{centering}
\vspace{-0.5cm} \includegraphics[width=12cm]{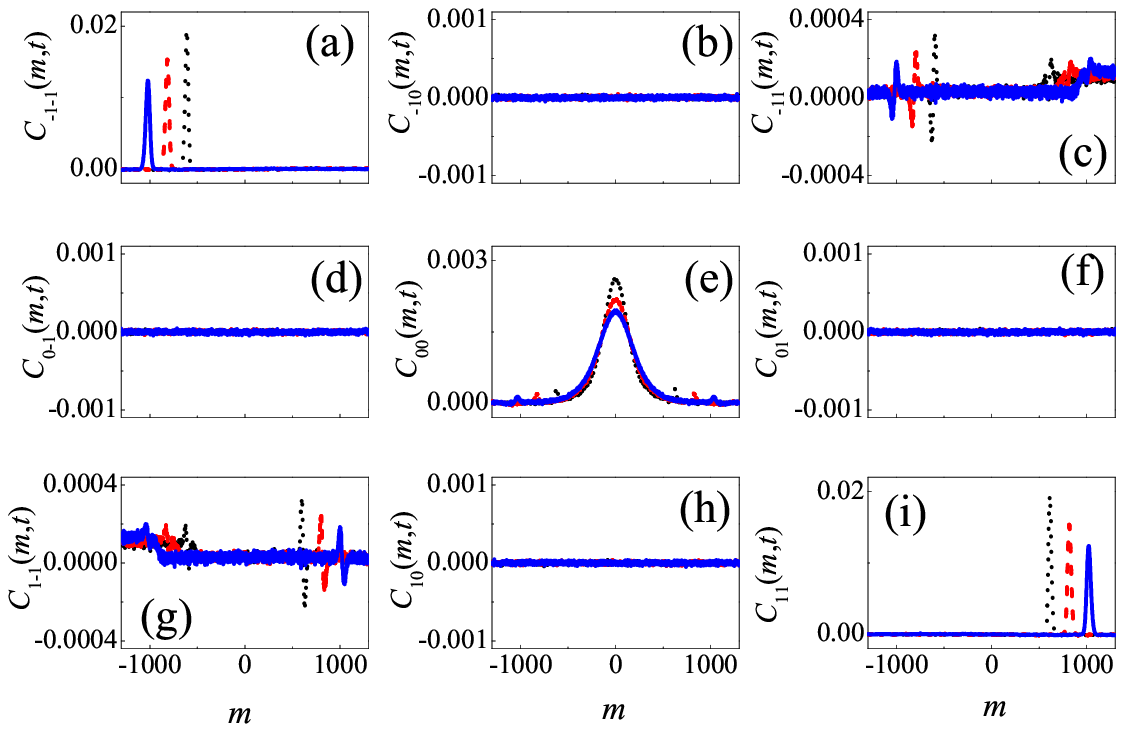} \vspace{-0.6cm}
\caption{\label{SFig1} The purely quartic anharmonic chain: the $3 \times 3$ correlation functions of heat and sound modes, for three long times (the same below): $t=600$ (dotted), $t=800$ (dashed), and $t=1000$ (solid), respectively.} \vspace{-0.6cm}
\end{centering}
\end{figure*}

\begin{figure}
\begin{centering}
\includegraphics[width=7cm]{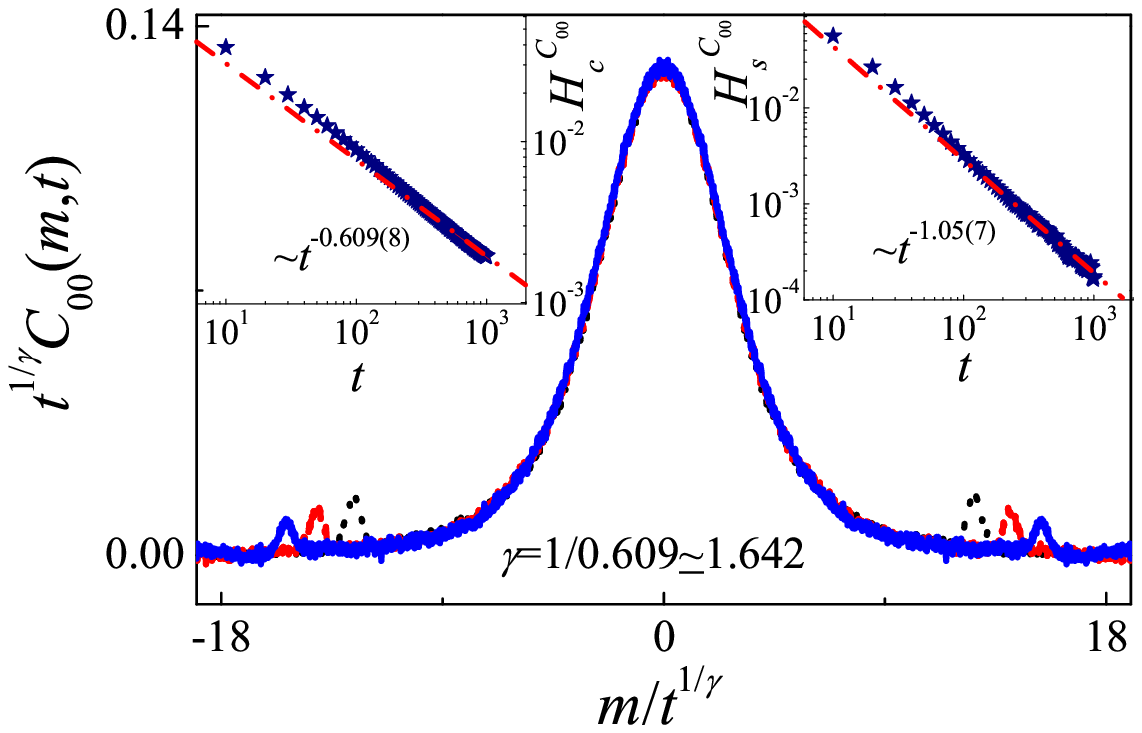} \vspace{-.5cm}
\caption{\label{SFig2} The purely quartic anharmonic chain: Rescaled $C_{00}(m,t)$ for three long times. The left (right) inset shows the height $H_c^{c_{00}}$ [$H_s^{c_{00}}$] of the central (side) peak of $C_{00}(m,t)$ vs $t$.} \vspace{-.3cm}
\end{centering}
\end{figure}

For the purely quartic anharmonic chain under temperature $T=0.5$ and pressure $\langle F \rangle=0$~\cite{S_Spohn2014},
\begin{equation} \label{Rmatrix}
R=
\left(
  \begin{array}{ccc}
    -1.02277 & -1 & 0 \\
    0 & 0 & 2.3094 \\
    -1.02277 & 1 & 0
  \end{array}
\right).
\end{equation}
We first estimate the $3 \times 3$ correlation functions of conserved field, then after a transformation we obtain the relevant normal modes correlations and show them in Fig.~\ref{SFig1}. As can be seen, the heat mode correlation $C_{00}(m,t)$ shows a similar profile as the estimate in Fig.~1 [see Fig.~\ref{SFig1}(e)]; the scaling of the central peak indicates an exponent $\gamma \simeq 1.642$ (see Fig.~\ref{SFig2}), deviation from the prediction $\gamma=3/2$~\cite{S_Spohn2014} as well; the scaling of the side peaks display a similar behavior as that shown in Fig.~1 (see the right inset of Fig.~\ref{SFig2}). In addition, it would be interesting to observe certain correlation between two sound modes, i.e., $C_{-11}(m,t)$ and $C_{1-1}(m,t)$ [see Figs.~\ref{SFig1}(c) and (g)]. Finally, it is hard to detect any information from $C_{-10}(m,t)$; $C_{0-1}(m,t)$; $C_{01}(m,t)$, and $C_{10}(m,t)$ [see Figs.~\ref{SFig1}(b);(d);(f);(h)], and the cross-correlation between heat and sound modes can only be observed in the side peaks of $C_{00}(m,t)$ [see Fig.~\ref{SFig1}(e)]. This seems to support one of the key arguments of MCT~\cite{S_LeeDadswell2005}: the thermal transport at any (sufficiently low) frequency can be determined entirely by
the thermal transport and momentum transport at much higher frequencies, while the vice versa will not happen; since here the sound (heat) mode corresponds to the low (high) frequency component.

\begin{figure}
\begin{centering}
\vspace{-.1cm}\includegraphics[width=7cm]{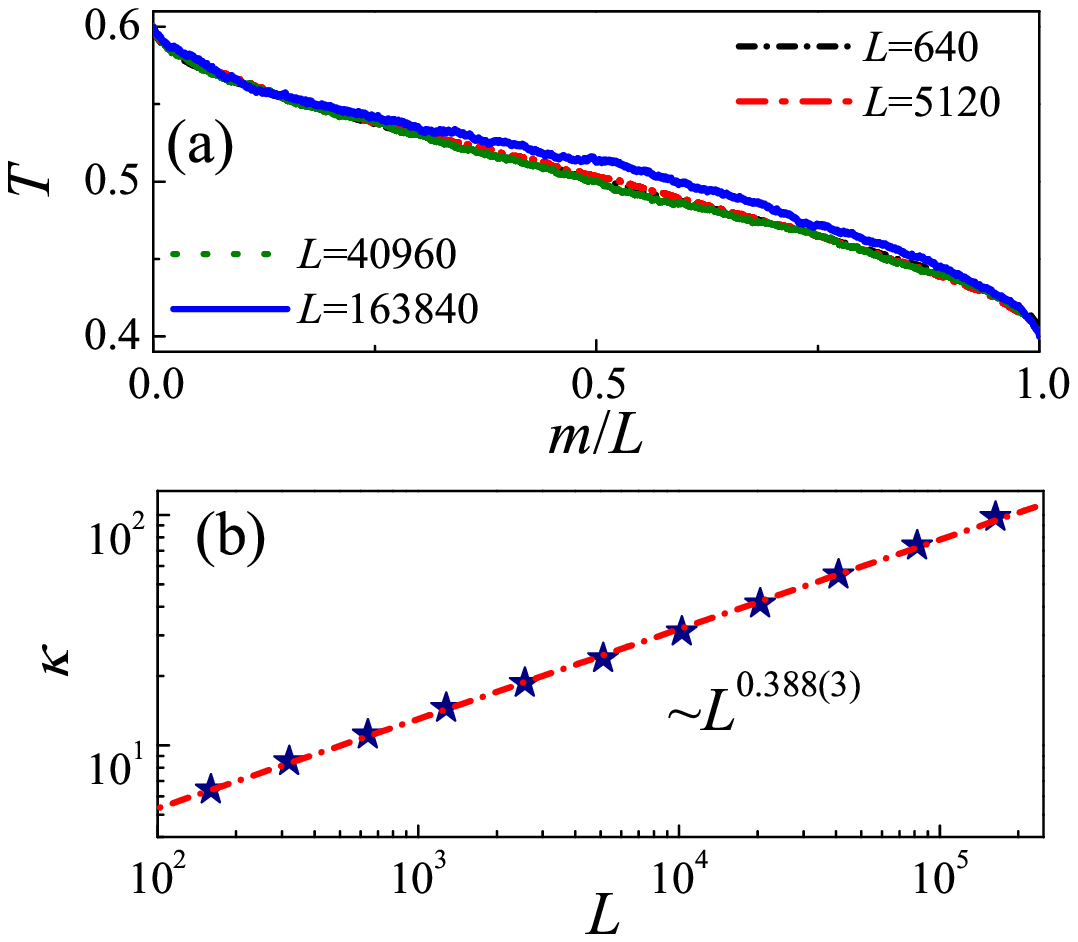} \vspace{-.4cm}
\caption{\label{SFig3} The purely quartic anharmonic chain: (a) Temperature profiles for four long $L$; (b) The measured $L$ dependence of $\kappa$.}
\vspace{-.3cm}
\end{centering}
\end{figure}

\emph{3. Estimate of $\alpha$.}---To numerically measure $\alpha$ for the purely quartic anharmonic chain, we adopt the direct nonequilibrium approach~\cite{S_Lepri1997}. It connects N\'{o}se-Hoover thermal reservoirs~\cite{S_Book} to the end particles of the chain with temperatures $T_+=0.6$ and $T_-=0.4$, and thus the averaged temperature along the chain is located at $T=0.5$. After long times (at least up to $4 \times 10^8$) evolution and when a steady state has been obtained, we then check the temperature profiles. As expected, in Fig.~\ref{SFig3}(a) all the temperature profiles curves for different $L$ indicate a proper scaling behavior. With this facility, the thermal conductivity $\kappa$ then can be measured by $\kappa=J L /(T_+-T_-)$. Fig.~\ref{SFig3}(b) shows the system-size dependence of $\kappa$ up to $L=163840$. This long $L$ is used to obtain the asymptotic behavior of the power-law divergence. The fitting, which is validated nearly perfectly for all the considered $L$ (from $L=160$ to $L=163840$, four orders of size), shows $\alpha=0.388 \pm 0.003$. This scaling seems to support $\gamma=\psi$ if the formula $\gamma=2-\alpha$ is valid.

\emph{4. Certain negative correlation.}---We find that the distorted side peaks shown in $\rho_Q(m,t)$ for the cubic-plus-quartic anharmonic chain ($A>0$) is related to certain negative correlation between heat and sound modes. For this we take $A=0.25$ for example. In the particle label description, under temperature $T=0.5$, the measured pressure is $\langle F \rangle \simeq -0.3262$, nearly twice that ($\langle F \rangle \simeq -0.176$) shown in the bin description (since the bin size is $2$). Applying this pressure into Eq.~(8.13) of~\cite{Spohn2014}, we obtain:
\begin{equation} \label{Rmatrix2}
R=
\left(
  \begin{array}{ccc}
    -1.01608 & -1 & -0.04184 \\
    -0.07542 & 0 & 2.31211 \\
    -1.01608 & 1 & -0.04184
  \end{array}
\right).
\end{equation}
\begin{figure*}
\begin{centering}
\vspace{-0.5cm} \includegraphics[width=12cm]{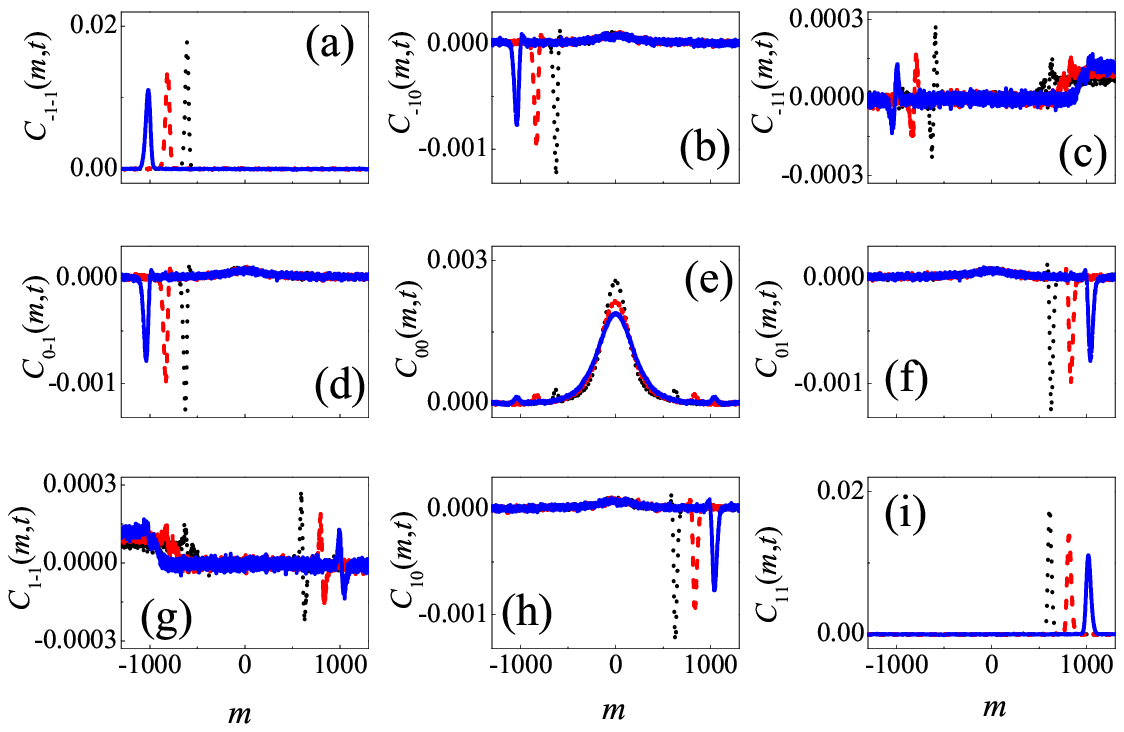} \vspace{-0.6cm}
\caption{\label{SFig4} The cubic-plus-quartic anharmonic chain ($A=0.25$): the same plot as Fig.~\ref{SFig1}.} \vspace{-0.6cm}
\end{centering}
\end{figure*}
With this and following similar NFHT scheme as above, we are able to derive the corresponding normal modes correlations. Fig.~\ref{SFig4} depicts such a result. In most cases, it shows similar profiles as Fig.~\ref{SFig1}, whereas for $C_{-10}(m,t)$; $C_{0-1}(m,t)$; $C_{01}(m,t)$, and $C_{10}(m,t)$, additional information now can be detected [see Figs.~\ref{SFig4}(b);(d);(f);(h)]. This seems to support that, instead of the fact: the energy (thermal) transport depends only on momentum transport in the particular purely quartic anharmonic chain, here, for the cubic-plus-quartic anharmonic chain, both the coupling between energy and momentum transport should be taken into account, which is in agreement with the theoretical argument of MCT~\cite{S_LeeDadswell2005}.
In particular, at the location of the side peak of the heat mode correlation, certain negative correlation of $C_{-10}(m,t)$; $C_{0-1}(m,t)$; $C_{01}(m,t)$, and $C_{10}(m,t)$ can be clearly identified. This last point just corresponds to the distortion of the side peaks of $\rho_Q(m,t)$ observed in Fig.~3. Since such distortion is expected to affect the scaling of $\rho_Q(m,t)$'s central peak, the new observed exponent $\gamma \simeq 1.580$ is likely to appear.


\begin{thebibliography}{99}
\bibitem{Note-r} In view of this similarity, we denote them as $\gamma$ as well and emphasize that, even though such scaling exponents are realized by different specific models, the relevant physics would be the same described by NFHT, see also the argument in~\cite{Spohn2014}.
\end{thebibliography}
\end{document}